\documentclass[aps,twocolumn,amsmath,preprintnumbers,amsfonts,superscriptaddress]{revtex4}

\usepackage{graphicx}
\def\beq{\begin{equation}}
\def\eeq{\end{equation}}
\def\beqa{\begin{eqnarray}}
\def\eeqa{\end{eqnarray}}

\def\za{\alpha}
\def\zb{\beta}
\def\ssc{\scriptscriptstyle}
\def\lsim{\mathrel{\raise.3ex\hbox{$<$\kern-.75em\lower1ex\hbox{$\sim$}}} }
\def\gsim{\mathrel{\raise.3ex\hbox{$>$\kern-.75em\lower1ex\hbox{$\sim$}}} }

\begin{document}
\draft
\preprint{\tighten{\vbox{\hbox{NCU-HEP-k034}
\hbox{Jun 2009}
\hbox{rev. Sep 2009}\hbox{ed. Nov 2009}
}}}

\title{Holomorphic Supersymmetric Nambu--Jona-Lasinio Model \\
with Application to Dynamical Electroweak Symmetry Breaking
}
\author{\bf Dong-Won Jung
}
\affiliation{
Department of Physics and Center for Mathematics and Theoretical Physics,
National Central University,~Chung-li,~TAIWAN 32054.  \\
        }
\author{\bf Otto C. W. Kong 
}
\affiliation{
Department of Physics and Center for Mathematics and Theoretical Physics,
National Central University,~Chung-li,~TAIWAN 32054.  \\
        }
\author{\bf Jae Sik Lee 
}
\affiliation{
Physics Division, National Center for Theoretical Sciences,~Hsinchu,~TAIWAN 300. \\
 }

\begin{abstract}
Based on our idea of an alternative supersymmetrization of the
Nambu--Jona-Lasinio model for dynamical symmetry breaking, we analyze
the resulting new model with a holomorphic dimension-five operator in the
superpotential. The approach provides a new direction for modeling dynamical
symmetry breaking in a supersymmetric setting. In particular, we adopt
the idea to formulate a model that gives rise to the Minimal Supersymmetric
Standard Model as the low energy effective theory with both Higgs
superfields as composites. A renormalization group analysis is performed
to establish the phenomenological viability of the scenario, with
admissible background scale that could go down to the TeV scale. We give
the Higgs mass range predicted.
\end{abstract}

\maketitle

\section{Introduction}
The Nambu--Jona-Lasinio (NJL) model~\cite{NJL} is a classic
model on dynamical symmetry breaking. A dimension-six operator of
four-fermion interaction is used to induce a bi-fermion vacuum
condensate. The bi-fermion configuration behaves as a scalar
composite; that is, the effective Higgs field responsible for symmetry
breaking and Dirac fermion mass. The possibility of a different
mechanism for the electroweak symmetry breaking
is a very interesting and inspiring option.

For the NJL model to give the vacuum condensate, the four-fermion
interaction needs to have a strong enough coupling. With the effective Higgs
multiplet introduced as an auxiliary field, the strong four-fermion coupling
translates into a large Yukawa coupling realizing plausibly the heavy top quark
phenomenologically. Such a top condensate model was constructed in
1989~\cite{top1,top2,top}. We refer readers to Ref.~\cite{rev} for a comprehensive
review of the details of the model and other related topics.

The investigation of supersymmetric version of the NJL model was
started in the 1980s. As a direct supersymmetrization, a dimension-six operator
that has a four-fermion interaction in its $D$ term was introduced ~\cite{BLE}.
It was realized that nontrivial vacuum is not possible due to supersymmetric
cancelation, unless soft supersymmetry (SUSY) breaking is incorporated. The
scheme introduces two new chiral superfields in the low energy effective
field theory with asymmetric roles. The approach leads to the most
popular candidate theory beyond the Standard Model (SM) --- the Minimal
Supersymmetric Standard Model (MSSM) as the low energy effective field
theory, with interesting relations among some of the model
parameters ~\cite{CLB,CCWBS}.
Compared to the non-SUSY model, it improves on
or eliminates a fine-tuning problem on the four-fermion coupling
while allowing the lower top mass. Incorporating
the NJL mechanism into the MSSM has the advantage of leaving
the model superfield spectrum with only the part that is
strongly constrained by the gauge symmetries, without the
otherwise unconstrained vectorlike pair of Higgs superfields,
besides enriching the naive Higgs mechanism with a conceptually
more appealing dynamical structure.
Phenomenological study of the model scenario was implemented with the infrared
quasi-fixed-point (IQFP) solution for the top quark mass~\cite{CCWBS}. In
a similar spirit, Ref.~\cite{FMK} presents an
IQFP determination of all third generation fermion masses in the MSSM.
That IQFP scenario should correspond to having both Higgs superfields as
composites, which is not compatible with the conventional supersymmetric
Nambu--Jona-Lasinio (SNJL) model.

Here, we propose a holomorphic variant of the SNJL model.
Instead of a dimension-six operator,
we consider a holomorphic dimension-five one in the superpotential. We
illustrate how the scenario may be used for dynamical symmetry
breaking. While the dimension-five term does not contain the
four-fermion interaction as a component, the model does have other features that
resemble the non-SUSY NJL model more closely compared to the old SNJL
model. We discuss how a model of such kind can give rise to
the MSSM as the low energy effective field theory realizing the IQFP
solution of Ref.~\cite{FMK}.  To fully accommodate the masses and mixing of
the quarks in the MSSM, we find the kind of dimension-five four-superfield
interactions actually cannot be avoided even in the old SNJL model.
However, none of the dimension-five terms plays a role in inducing
symmetry breaking in that case.

Inspired by the holomorphic model, we implemented a renormalization group (RG)
analysis of the MSSM numerically, looking for compatibility with the model
scenario. One would think that the current top mass of $171.3\pm1.6$~GeV
is very difficult to be accommodated by a NJL model. It is not the case
for our holomorphic SNJL model scenario. With a large enough value for
$\tan\!\beta$, fitting the experimental quark masses presents little problem.
And somewhat to our surprise, we realize that the bottom quark Yukawa
coupling plays a more important role than the top Yukawa, and the composite
scale could be very low. With a simple calculation, the Higgs mass turns up
close to the current bounds. Note that, phenomenologically, fitting the
current top mass with the MSSM from the old SNJL model actually pushes the
$\tan\!\beta$ value into a narrow window between $0.5$ to $1.5$ which is
essentially excluded by the LEP result. Our holomorphic version
as the first complete model for the MSSM, however, is
phenomenologically viable.

We conclude that the holomorphic SNJL model
idea provides an interesting alternative to the construction of dynamical
symmetry breaking models and the case for such a model as what is behind the
MSSM and the hence the phenomenology at the LHC scale is worth a
more serious investigation.

\section{The Holomorphic SNJL Model}
Consider the following model Lagrangian :
\small
\beq \label{eq:HSNJL}
{\cal L}=\int \!d^4\theta \left[\bar{\Phi}_+ {\Phi}_+
+ \bar{\Phi}_- {\Phi}_- \right]
- 
\int \!d^2\theta
\frac{G}{2} \,{\Phi}_+
{\Phi}_- {\Phi}_+ {\Phi}_-
+ h.c. 
\;.
\eeq \normalsize
Besides the kinetic terms, we have introduced a four-superfield interaction
term, but it is in the superpotential hence of dimension-five only. In the
simplest version of the model, we need only one auxiliary
chiral superfield ${\Phi}_0$  to rewrite the Lagrangian (\ref{eq:HSNJL})
with a Yukawa coupling. Explicitly,
we consider
\small
\beqa
{\cal L} &=&\int\! d^4\theta \left[ (\bar{\Phi}_+ {\Phi}_+
+ \bar{\Phi}_- {\Phi}_-)(1-m^2 \theta^2\bar{\theta}^2) \right]
\nonumber \\
&& + \left\{ \int\! d^2\theta \left[ \frac{1}{2}  ( \sqrt{\mu}{\Phi}_0
+\sqrt{G}  {\Phi}_+ {\Phi}_-) ( \sqrt{\mu}{\Phi}_0
+\sqrt{G}  {\Phi}_+ {\Phi}_-)
\right.\right. \nonumber \\ && \left.\left.
- \frac{G}{2} {\Phi}_+
{\Phi}_- {\Phi}_+ {\Phi}_- \right] + h.c. \right\}
\nonumber \\
&=&  \int\! d^4\theta \left[ (\bar{\Phi}_+ {\Phi}_+
+ \bar{\Phi}_- {\Phi}_-)(1-m^2 \theta^2\bar{\theta}^2) \right]
 \nonumber \\ &&
+\left\{\int\! d^2\theta \left[ \frac{\mu}{2} {\Phi}_0^2
+ \sqrt{\mu G}{\Phi}_0 {\Phi}_+ {\Phi}_- \right] + h.c.\right\}\;,
\label{L5e}
\eeqa \normalsize
where we have again put in a soft SUSY breaking term.
Without the latter, the last line in the equation is an exact
supersymmetrization of the NJL counterpart.
Note that the superpotential reduces to
\beq
W  = - \frac{\mu}{2} {\Phi}_0 \left[ {\Phi}_0 + 2\sqrt{G/\mu}
{\Phi}_+ {\Phi}_- \right]\;,
\eeq
to be compared against $W= -\mu{\Phi}_2 ({\Phi}_1 + g {\Phi}_+ {\Phi}_-)$
for the old SNJL model. The equation of motion for ${\Phi}_0$ yields
${\Phi}_0 = -\sqrt{G/\mu}\,{\Phi}_+ {\Phi}_-$, {\it i.e.} ${\Phi}_0$
as a composite of two chiral superfields. Simultaneously, ${\Phi}_0$
also takes the role of the effective Higgs superfield, supersymmetrizing the
composite scalar field  of the original NJL model. The
mathematical structure of our Lagrangian thus resembles the latter
more closely compared to that of the old SNJL model. In the latter case,
${\Phi}_1$ is the composite while ${\Phi}_2$ plays the Higgs.

To look at the physics of the model and its possible symmetry breaking feature,
we follow a simple approach discussed in Ref.~\cite{CLB} for a gauged version of the
old SNJL model. The approach is also discussed in Ref.~\cite{rev} for the NJL model,
where it is explicitly shown to give the same result as the gap equation analysis.
For this purpose, we calculate the effective two point function for the auxiliary
superfield $\Phi_0$ of the Lagrangian in Eq.~(\ref{L5e}). From the one-loop supergraph
diagram with Yukawa vertices, we have,
in the presence the extra soft SUSY breaking $m^2$ terms,
{\small
\begin{eqnarray}
\Gamma_{eff}\!\! &\simeq& \!\!\int d^4x d^4\theta \left(\frac{y^2}{16 \pi^2}
\log \left[\frac{\Lambda^2}{\mu_Q^2} \right] \right)
\bar{\Phi}_0 \Phi_0 \left[1+ 2 m^2\theta^2 \bar{\theta}^2 \right]\nonumber \\
&\equiv& \int d^4x d^4\theta ~Z_0  \bar{\Phi}_0  \Phi_0
\left[1+ 2 m^2\theta^2 \bar{\theta}^2 \right] \;.
\end{eqnarray}
}
Note that SUSY breaking mass induced for the effective canonical Higgs superfield
$\sqrt{Z_0}\Phi_0$ is tachyonic, suggesting the possibility of radiatively
induced symmetry breaking.
Actually, the one Higgs case is of very limited interest.
To have the ${\Phi}_0^2$ term in the Lagrangian, ${\Phi}_0$ has to be in a real
representation of the model symmetry. The electroweak doublet needed
for the  SM symmetry breaking, for instance, cannot be modeled directly.


\section{Towards the MSSM}
The supersymmetric SM requires two Higgs superfields, instead of one.
Consider a Lagrangian of four chiral superfields (actual the three
third-generation quark superfields) with soft SUSY breaking masses and the
following superpotential
\beq \label{W}
W = G \, \varepsilon_{\!\za\zb}{Q}^{\za a}_{\!\ssc 3} {U}^{c\, a}_{\!\ssc 3}
 {Q}^{\zb b}_{\!\ssc 3}{{D}^{\!c \,b}_{\!\ssc 3}} (1+A\theta^2) \;,
\eeq
where we use standard notation for quark doublet and singlet superfields;
$\za,\zb$ are $SU(2)$ indices and $a,b$ are color indices. Two Higgs superfields
are introduced to rewrite the superpotential, with the $SU(2)$ and color
indices suppressed, as the equivalent
\beqa
W
-\mu ({H}_d
 - \lambda_t {Q}_{\!\ssc 3} {U}^{c}_{\!\ssc 3} ) ({H}_u - \lambda_b {Q}_{\!\ssc 3}{D}^{\!c}_{\!\ssc 3} )(1+A\theta^2)
&&
\nonumber \\
=(-\mu {H}_d {H}_u + y_t {Q}_{\!\ssc 3} {H}_u {U}^{c}_{\!\ssc 3}+ y_b {H}_d {Q}_{\!\ssc 3}{D}^{\!c}_{\!\ssc 3} ) (1+A\theta^2)\;,
&&
\eeqa
where $\mu \lambda_t =y_t$, $\mu \lambda_b =y_b$, $\mu \lambda_t \lambda_b =G$.
The equation of motion for ${H}_u$ gives
${H}_d= \lambda_t {Q}_{\!\ssc 3} {U}^{c}_{\!\ssc 3}$ while that for ${H}_d$ yields
${H}_u = \lambda_b {Q}_{\!\ssc 3}{D}^{\!c}_{\!\ssc 3}$. Note that the SUSY breaking part
with parameter $A$ also gives the $B$ term. Promoting the quark-superfield kinetic
terms to full gauge kinetic terms and adding the pure gauge superfield terms,
one arrives at a Lagrangian for third generation quark masses similar to the one
considered in Ref.~\cite{CCWBS} for the old SNJL model. In our holomorphic model,
however, both ${H}_u$ and  ${H}_d$ come as quark-superfield composites, with
gauge kinetic terms expected to be generated at low energy through the Yukawa
couplings~\cite{CLB}.

A full superpotential for the MSSM Lagrangian with the two
effective Higgs superfields can be given by
{\beq
W = G_{i\!j\!k\!h} \, {Q}_i {U}^{c}_j {Q}_k{D}^c_h (1+A\theta^2)
+ G_{i\!j}^e \, {Q}_{\!\ssc 3} {U}^{c}_{\!\ssc 3} {L}_i{E}^c_j (1+A\theta^2) \;,
\eeq}
with the usual family indices. We assume only the coupling $G_{\ssc 3333}$ is strong
enough to drive the dynamical symmetry breaking as described above.
The $A$ parameter does not have to be universal, but we do have
the MSSM $A_t$ and $A_b$ and the usual $B$-parameter originate from a single soft
SUSY breaking parameter for the  $G_{\ssc 3333}$ term, as in Eq.~({\ref{W}) above.
The four-superfield terms with couplings $G_{ij{\ssc 33}}$,  $G_{{\ssc 33}kh}$
and $G_{ij}^e$ give rise to the rest of the up-type quark, down-type quark,
and charged lepton Yukawa interactions, respectively. The remaining
$G_{i\!j\!k\!h}$ terms are not needed. Neither does it hurt to have the extra
$G_{i\!j}^e$ type terms so as to include all terms admissible by the gauge symmetry.

At this point, it is interesting to go back and look at how the old SNJL model
get to the MSSM itself. While some careful numerical studies have been
performed (see, for example, Ref.~\cite{CCWBS}), apparently, the full
Lagrangian has not been explicitly given at the level before introducing the
auxiliary superfields. The original dimension-six term gives the composite
$H_d \sim {Q}_{\!\ssc 3} {U}^{c}_{\!\ssc 3}$ with the other auxiliary
$H_u$ to give up-type quark Yukawa terms.
Duplicating the structure to generate the down-sector quark Yukawa terms from
$\bar{Q}\bar{D}^cQD^c$ operators (and charged lepton part from
$\bar{L}\bar{E}^cLE^c$ terms) will introduce a further pair(s) of Higgs superfields.
On the other hand, sticking to using $H_d \sim {Q}_{\!\ssc 3} {U}^{c}_{\!\ssc 3}$
as the Higgs superfield for
the down-sector quark and charged lepton Yukawa terms implies exactly that
they come from the kind of holomorphic dimension-five operators introduced here.
For example, $\int\! d^2\theta Y^e H_d L E^c \longleftarrow \int\! d^2\theta
\left(-gY^e {Q}_{\!\ssc 3} {U}^{c}_{\!\ssc 3} L E^c\right)$.

\section{renormalization group analysis}
To see if the holomorphic SNJL model scenario discussed above is compatible with
the low energy phenomenology and known experimental constraints, we perform a
renormalization group (RG) analysis and report the first results here, leaving
most of the details to a forthcoming paper \cite{038}. Here, we focus only on
the most important parameters, the top and bottom quark Yukawa couplings. We
also take a look at the Higgs mass predicted.

\begin{figure}[b]
\begin{center}
\includegraphics[angle=270, scale=0.32]{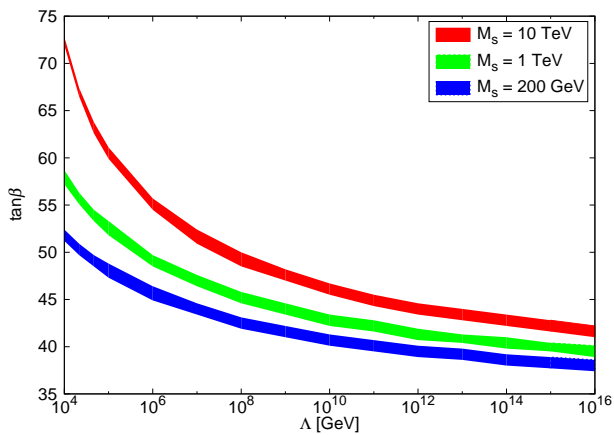}
\caption{Solution window for $\tan\!\beta$ vs $\Lambda_b$}
\end{center}
\label{fig:tb}
\end{figure}

\begin{figure}[h]
\begin{center}
\includegraphics[angle=270, scale=0.32]{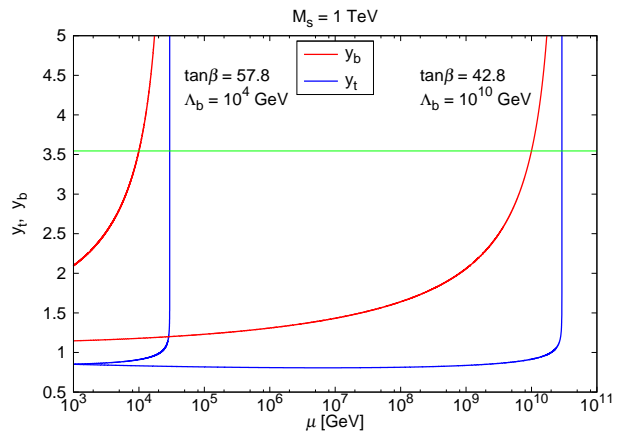}
\caption{Illustration of $y_b$ and $y_t$ runnings for a couple of cases.}
\end{center}
\label{fig:run}
\end{figure}

The IQFP scenario had been the focus in most of the RG
analysis related to the NJL models in the literature. We first note that while
the scenario has its appeal, especially when one aims at getting a prediction
for the top mass, it is not strictly required by the NJL model mechanism.
The latter requires only for the related Yukawa couplings, those for both the
top and the bottom quark in our case, to blow up approaching the background
scale $\Lambda$. As our analysis is based on one-loop RGEs, we may not reliably
trace the running beyond the scale where the couplings pass the
perturbative limit~\cite{top}. Denote the scales by $\Lambda_t$ and $\Lambda_b$
({\it i.e.} where $y^2/4\pi=1$), respectively. We look for admissible
cases of $\Lambda_t$ and $\Lambda_b$ values with the now precisely determined top
and bottom masses $\left[m_b=4.20^{+0.17}_{-0.07}\,\rm{GeV}\right]$ implemented.
Note that for large value of $\tan\!\beta$, the bottom Yukawa $y_b$ is big.
We find that there is always a window of $\tan\!\beta$ value giving admissible
solution, for any $\Lambda_b$ we take (with SUSY scale $M_S$ from 200~GeV to
10~TeV), as given in Fig.~1. In Fig.~2, we show the $y_t$ and $y_b$ runnings
for a couple of typical cases.  Note that $y_b$  plays the most important
role, reaching the perturbative limit before $y_t$. We have
$\Lambda_t \sim 3\Lambda_b$, with the one-loop RG showing divergent behavior
for both $y_t$ and $y_b$ almost right beyond $\Lambda_t$. It is also
important to note that we have included a SUSY threshold correction $\epsilon_b$
[{\it cf.~}$(1+\epsilon_b \tan\!\beta)={\sqrt2}m_b/(y_b\,v\,\cos\!\beta)$] of value
$-0.01$ in the running of $y_b$. The negative value  is important. For $\epsilon_b>0$,
solution is possble only with uncomfortably large $\tan\!\beta$.
The exact value of $\epsilon_b$ depends
on the SUSY spectrum. For the simple case of a degenerate spectrum, we have
$|\epsilon_b|=\alpha_s/3\pi \sim 0.01$\cite{eb}. For the first
result here, we take the ballpark value. Looking at the RG running only,
the yet to be determined value of $\tan\!\beta$ means that it can be chosen to
fix the low energy $y_b$ input value to yield almost any $\Lambda_b$ value.
The negative $\epsilon_b$ help to reduce somewhat the $\tan\!\beta$
value thus taken, keeping $y_t$ not too small to fit in the RG picture. That
explains the main feature behind our result.
If one takes a careful inspection of the admissible solution plot, one will see
that for a large enough $\Lambda_b$, the $\tan\!\beta$ solution window loses
sensitivity to further increase in $\Lambda_b$. This is what is corresponding
to the IQFP solution, as a variation of $\Lambda_b$
translates into a variation of $y_b$ at a fixed $\Lambda$.

For the determination of the Higgs mass, we follow the approach of
Ref.\cite{CCWBS}. The MSSM Higgs potential parameters for the quartic terms
are to be fixed at the $M_S$ scale by the gauge couplings, and then run
towards the lower energy scale with the appropriate RG equations. Together with
the $\tan\!\beta$ value, the Higgs potential, assumed to give the right
electroweak symmetry breaking, is left with only one free parameter here
taken as $M_A$, the pseudoscalar mass. We determine the lightest Higgs mass
as a function of $M_S$ for $M_A>100$~GeV. The value loses sensitivity to
$M_A$ as the latter get bigger. In fact, for large $M_A$, a SM Higgs potential
should rather be used with the only parameter fixed at $M_S$. We confirmed
a good agreement of the result for the case. We present in Fig.~3 the result
for the lightest Higgs mass. For $M_S<1.5$~TeV, it is on the low side compared
with the 114~GeV SM Higgs search limit, but with admissible values as
MSSM Higgs \cite{hm}. The result has little sensitivity to $\Lambda_b$.
We conclude that the Higgs mass we have is generally admissible,
while further studies explicitly tuned to various parameter space region are
needed, better to be done in conjunction with analyses of the full SUSY
spectrum. It is  plausible that the Higgs mass will be increased after
further corrections are taken into account.

\begin{figure}[t]
\begin{center}
\includegraphics[scale=0.32, angle=270]{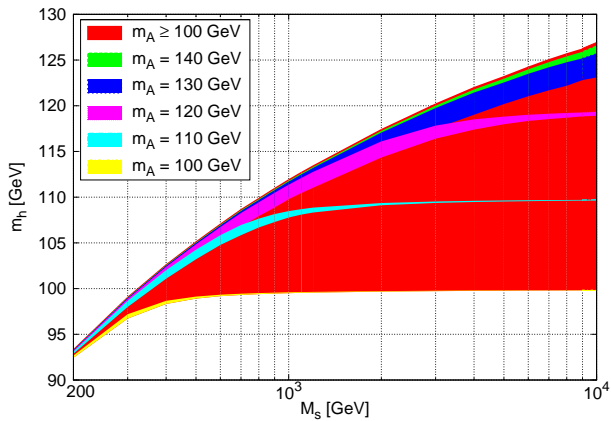}
\caption{Prediction for the lightest Higgs mass.}
\end{center}
\label{fig:mh}
\end{figure}

\section{Final Remarks}
We have explored above the idea of having a holomorphic SNJL model.
The structure is unconventional and provocative in the sense that it actually
requires a bi-scalar vacuum condensate to give symmetry breaking and
Dirac masses. The bi-fermion composite/condensate setting was inspired
by the Cooper pair and the BCS theory of superconductivity, and quite
well explored with various approaches~\cite{rev}. Our discussion for
the bi-scalar condensate scenario in this paper is  short of
establishing it on a similar ground.
The scenario, however, may provide a new direction for modeling
dynamical symmetry breaking in a supersymmetric setting.

We have discussed a holomorphic SNJL model with two Higgs superfields
giving complete MSSM Lagrangian as the low energy effective theory.
The latter may realize the IQFP solution for third family fermion masses as
studied in Ref.~\cite{FMK}, even for $m_t=171.3\pm 1.6$~GeV.
More interestingly, a much lower background scale $\Lambda$ for the
SNJL model, while deviating from the fixed point picture, still gives admissible
solution for a somewhat $\Lambda$-sensitive window of $\tan\!\beta$. Even
$\Lambda$ of the TeV order may be admissible. For $M_S<1.5$~TeV,
the Higgs mass is on the low side, but compatible with
the MSSM Higgs search limits.
The model also prefers $\epsilon_b$ to be negative.

The RG analysis presented here includes only the first results. There are
other interesting parts to be investigated, for example, the behavior and
preferred values of the soft SUSY breaking parameters. Detailed comparison
with the SUSY-top condensate scenarios should also be studied. More
results will be given in a forthcoming publication~\cite{038}.

\acknowledgements
D.-W.J. and O.K. is partially supported by research grant number
96-2112-M-008-007-MY3, and D.-W.J. is further supported by grant number
 098-2811-M-008-032 of the NSC of Taiwan.

{{\it Note Added:-}
After we posted the first version of this paper, we were kindly informed by
Kobayashi and Terao about their study on a
supersymmetric QCD model with a complicated Higgs sector\cite{KT},
with Higgs superfields essentially in a real representation. Seeking to eliminate
the Higgs superfields as auxiliary composites, the author arrived at the
dimension-five operator to be taken as the source of the
Higgs superfields and the origin of the (dynamical) symmetry breaking.
This is essentially the holomorphic SNJL model, which we rediscovered and analyzed
from the basic supersymmetrizing perspective. It is easy to see from our discussion
above that assuming essentially only one Higgs superfield multiplet cannot fit into
the conventional SNJL model with the dimension-six operator, but works with
the holomorphic model.
}

\end{document}